# Toward Mass-Production of Transition Metal Dichalcogenide Solar Cells: Scalable Growth of Photovoltaic-Grade Multilayer WSe$_2$ by Tungsten Selenization

Kathryn M. Neilson,[1†] Sarallah Hamtaei,[2,3,4†] Koosha Nassiri Nazif,[1†] Joshua M. Carr,[5] Sepideh Rahimisheikh,[6] Frederick U. Nitta,[1,7] Guy Brammertz,[2,3,4] Jeffrey L. Blackburn,[8] Joke Hadermann,[6] Krishna C. Saraswat,[1,7] Obadiah G. Reid,[8,9] Bart Vermang,[2,3,4] Alwin Daus,[1,10] & Eric Pop[1,7,11]

[1]Dept. of Electrical Engineering, Stanford Univ., Stanford, CA 94305, USA

[2]Hasselt Univ., imo-imomec, Martelarenlaan 42, Hasselt 3500, Belgium

[3]Imec, imo-imomec, Thor Park 8320, Genk 3600, Belgium

[4]EnergyVille, imo-imomec, Thor Park 8320, Genk 3600, Belgium

[5]Univ. Colorado Boulder, Materials Science & Engineering Program, Boulder CO 80303, USA

[6]Univ. Antwerp, Electron Microscopy for Materials Science (EMAT), Groenenborgerlaan 171, Antwerpen 2020, Belgium

[7]Dept. of Materials Science & Engineering, Stanford Univ., Stanford, CA 94305, USA

[8]National Renewable Energy Laboratory, Chemistry and Nanoscience Center, Golden CO 80401, USA

[9]University of Colorado Boulder, Renewable and Sustainable Energy Institute, Boulder CO 80303, USA

[10]Sensors Laboratory, Dept. Microsystems Engineering (IMTEK), Univ. Freiburg, Freiburg, Germany

[11]Precourt Institute for Energy, Stanford Univ., Stanford, CA 94305, USA

[†]These authors contributed equally. *Corresponding author email: epop@stanford.edu

**Semiconducting transition metal dichalcogenides (TMDs) are promising for high-specific-power photovoltaics due to desirable band gaps, high absorption coefficients, and ideally dangling-bond-free surfaces. Despite their potential, the majority of TMD solar cells are fabricated in a non-scalable fashion using exfoliated materials due to the absence of high-quality, large-area, multilayer TMDs. Here, we present the scalable, thickness-tunable synthesis of multilayer tungsten diselenide (WSe$_2$) films by selenizing pre-patterned tungsten with either solid source selenium or H$_2$Se precursors, which leads to smooth, wafer-scale WSe$_2$ films with a layered van der Waals structure. The films have charge carrier lifetimes up to 144 ns, over 14× higher than large-area TMD films previously demonstrated. Such high carrier lifetimes correspond to power conversion efficiency of ~22% and specific power of ~64 W g$^{-1}$ in a packaged solar cell, or ~3 W g$^{-1}$ in a fully-packaged solar module. This paves the way for the mass-production of high-efficiency multilayer WSe$_2$ solar cells at low cost.**

Semiconducting transition metal dichalcogenides (TMDs), e.g., MoS$_2$ and WSe$_2$, offer excellent electronic and optical properties for use in a variety of applications from nanoelectronics to photovoltaics. These include good charge carrier mobility in atomically thin (sub-1-nm) layers, ultrahigh optical absorption coefficients, near-ideal band gaps for solar energy harvesting, surfaces without dangling bonds, and facile integration on rigid and flexible substrates[1-4]. In particular, TMDs hold great promise in high-specific-power (i.e. high-power-per-weight) applications, e.g., energy harvesting in high-altitude drones and wearable electronics, where light weight and high power output are desired[5-7]. These emerging photovoltaic markets are growing at a rapid pace, potentially exceeding $100B by 2027[5].

Recent studies have shown that optimally-designed thin (<100 nm) TMD solar cells can in practice achieve ~25% power conversion efficiency (PCE) in a single-junction structure[2, 8] and ~32% PCE in a TMD-silicon tandem structure[9]. As a result, single-junction TMD solar cells can offer excellent specific power (power-per-weight) greater than 40 W g$^{-1}$, about 10× higher than established thin-film solar cell technologies cadmium telluride (CdTe), copper indium gallium selenide (CIGS), amorphous silicon (a-Si), and III-Vs[6, 7, 10]. In addition, while thin-film solar cell technologies face challenges such as high cost, low durability, and the use of rare, toxic materials, TMD solar cells provide a potentially low-cost, durable, eco-friendly, and bio-compatible solution which could reach large-scale adoption in the near future.

The majority of TMD synthesis research over the past decade has been focused on monolayer (< 1 nm) films, particularly for use in next-generation transistors. In contrast, high-quality TMD multilayers are most often obtained in a non-scalable fashion through the mechanical exfoliation of micron-scale flakes from bulk crystals grown by chemical vapor transport (CVT). Selenization (and/or sulfurization) of metallic precursor(s) could be employed as a low-cost method to synthesize TMD films at industrial scale, similar to CIGS[11]. In this approach, the substrate size is only limited by the furnace dimensions and can be easily scaled to larger areas with the use of industrial tools. However, previous studies on selenization synthesis of WSe$_2$ films have displayed suboptimal properties for photovoltaic applications, being either too thin[12], too rough[13], and/or suffering from morphological inhomogeneity[14]. The selenization process itself could also exceed 20 hours, which renders it unfavourable from an industrial perspective[15]. An ideal film growth would have a controllable thickness that is atomically smooth and oriented in the (001) direction, with sufficiently large grain sizes and good charge carrier lifetimes.

Herein, we demonstrate the synthesis of multilayer (15-30 nm) WSe$_2$, by selenization of tungsten on up to 150 mm wafers, in 1-2 hours. Such thicknesses correspond to an energy band gap of 1.2-1.3 eV, which are near-ideal for energy harvesting under both AM0 and AM1.5G spectra. The films exhibit high-quality characteristics, including a van der Waals (vdW) layered structure, smooth and uniform surfaces, grain sizes comparable to the film thickness, Hall mobilities up to 8 cm$^2$ V$^{-1}$ s$^{-1}$, and minority carrier lifetimes up to 144 ns; these minority carrier lifetimes are over 14× greater than previous reports of large-area TMD films, which were prepared by liquid exfoliation[16]. These WSe$_2$ films can provide efficiency up to 22.3% in an optimized solar cell design. More importantly, upon integration on ultrathin flexible substrates, such solar cells are calculated to provide specific power of ~64 W g$^{-1}$ in a packaged cell, over 10× higher than established thin-film solar cell technologies[10]. Fully packaged modules, which include considerations such as interconnect weight and encapsulation, show specific power of up to 3 W g$^{-1}$, approximately 5× higher than the record of 0.7 W g$^{-1}$, demonstrated in a multi-junction III-V-based solar module[5].

## WSe$_2$ Growth by Selenization of Tungsten

To prepare the WSe$_2$ film, we sputter 5-10 nm of W using a DC source (Kurt J. Lesker) at 50 W onto ~500 μm-thick silicon substrates at a pressure of 10 mTorr and an argon (Ar) flow rate of 50 sccm (**Fig. 1a**) to provide an optimal film porosity[14]. (See **Supplementary Fig. S1** for further discussion on the effect of sputtering pressure on growth quality.) These films can either be pre-patterned (via lithography and lift-off) or blanket-deposited. The W films are then selenized into multilayer (15-30 nm) WSe$_2$ by either solid source selenization (SS-Se) or by hydrogen selenide (H$_2$Se), as shown in **Fig. 1b-e**.

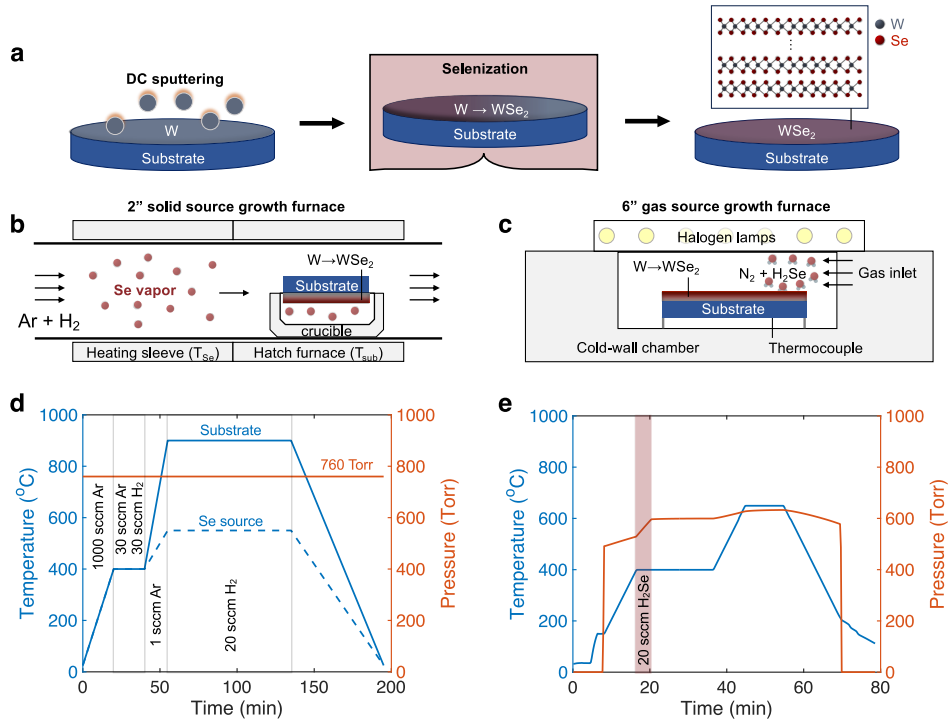

**Fig. 1: Scalable growth of WSe$_2$ by selenization of tungsten (W) using solid source selenium (SS-Se) or H$_2$Se precursors. a,** Selenization process flow. Furnace set-ups for **b,** SS-Se and **c,** H$_2$Se selenization processes. Optimized temperature and pressure vs. time for **d,** SS-Se and **e**, H$_2$Se processes. The dashed and solid blue lines in **d** correspond to $T_{Se}$ and $T_{sub}$ in **b**, respectively. H$_2$Se growth is done at a lower thermal budget (temperature) and pressure compared to SS-Se. Flow is continuously supplied for SS-Se (**d**) versus the finite introduction of H$_2$Se (**e**). Initial pressure changes in H$_2$Se growths are achieved by N$_2$ gas.

SS-selenization is carried out in a two-zone tube furnace (**Fig. 1b**) where Se pellets are placed in a crucible upstream, subsequently vaporized at elevated temperatures, then transported by carrier gas (Ar) to the main heating zone of the furnace containing the substrate. The substrate is placed face-down on a crucible to prevent sublimation of stoichiometric transition metal films at high temperatures[17], and both zones are heated to 400 °C for 20 minutes (**Fig. 1d**). Subsequently, the Se zone is heated to 550 °C and the substrate is heated to 900 °C for one hour. Given that the as-deposited W film may have natural surface oxidation before it is placed in the furnace, we flow additional $H_2$ gas to convert this surface into a more reactive $WO_{3-x}$ form[18]. The furnace is then left to naturally cool down to room temperature.

For the growth with $H_2Se$, a rapid thermal annealing tool is employed with nitrogen and $H_2Se$ as process gases (**Fig. 1c**). The sample is placed face-up on a silicon carbide (SiC)-coated graphite susceptor. The growth is done by a two-stage anneal (**Fig. 1e**); the chamber is filled with 488 Torr of $N_2$ base pressure, which is then heated up to 400 °C. The sample is then annealed for 20 minutes, with 75 Torr of diluted $H_2Se$ gradually fed into the chamber during the first 4 minutes, at 20 sccm. The chamber is subsequently heated to 650 °C, where the sample is held for 10 minutes before cooling to room temperature. Immediately after the two annealing stages (at 400 °C and 650 °C), the chamber pressure is 600 and 638 Torr, respectively. All heat-up and cool-down ramps are set to 0.5 °C/s.

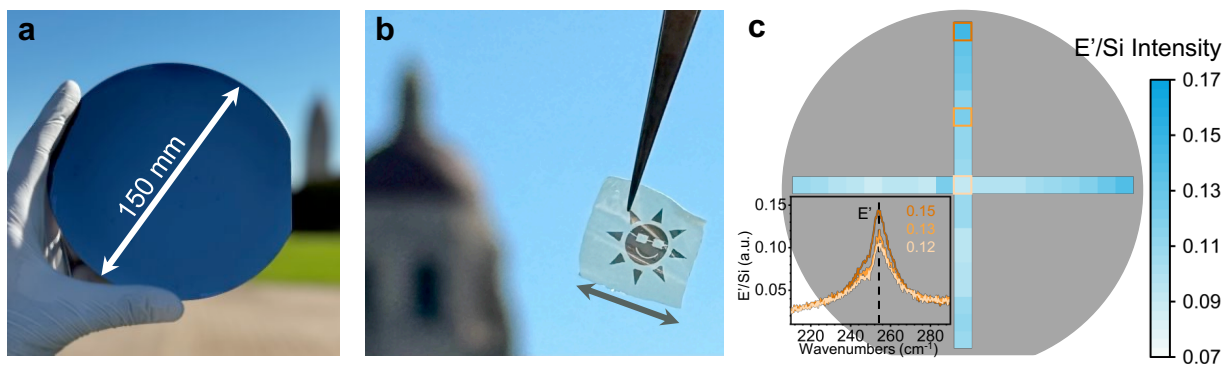

**Fig. 2. Versatility and scalability of the selenization growth method. a**, Photograph of a $WSe_2$ film uniformly grown by $H_2Se$ on a 150 mm wafer. The film size is only limited by the substrate and furnace dimensions. **b,** $WSe_2$ film obtained from SS-Se selenization of pre-patterned W, transferred onto a flexible lightweight polyimide substrate. (Selenization can be performed both on blanket-deposited and patterned W.) Due to their van der Waals nature, such selenized films can be easily transferred to other rigid or flexible substrates. **c,** Raman spectroscopy shows little change of the E' peak intensity (normalized by the Si peak) over two orthogonal wafer diameters, indicating wafer-scale $WSe_2$ film uniformity. Inset shows normalized Raman spectra of the $WSe_2$ at three spots marked by squares along the vertical scan.

**Figure 2** shows the versatility and scalability of the selenization growth method. Selenization of W leads to multilayer $WSe_2$ films whose thickness can be tuned by the amount of sputtered W. The substrate size is only limited by the furnace dimensions, which in our study is 50 mm for the solid-source and 150 mm for the gas-source selenization (**Fig. 2a**). Industrialized furnaces would enable growth on even larger substrates, i.e. 300 mm or greater. The $WSe_2$ films are highly uniform, with van der Waals structure, and can be easily transferred to other rigid or flexible substrates, e.g. to build flexible, high-specific-power solar cells[10]. For example, **Fig. 2b** shows pre-patterned WSe$_2$ film transferred onto a flexible, lightweight polyimide substrate, and **Fig. 2c** demonstrates the growth uniformity on a 150 mm wafer by displaying the highest intensity Raman peak (E') of WSe$_2$ normalized to the silicon substrate below.

## Material Characterization

Several characterization methods are used to assess the quality of the grown WSe$_2$ films. Initial scanning electron microscopy (SEM) images show smooth WSe$_2$ grown with H$_2$Se (**Fig. 3a**). Energy dispersive X-ray spectroscopy (EDS) is used to determine the Se to W ratio (**Fig. 3b-c**) of the films either as-grown on silicon or after transfer to PI. EDS spectra are taken under 5 kV acceleration voltage (**Fig. 3b**), to ensure data acquisition at the surface. The measurement is first done at the edge of the chip, as the smoothness of the films makes it difficult to focus near the center, followed by two more measurements at randomly chosen and wide apart locations.

**Supplementary Table 1** summarizes the average and coefficient of variation between these three measurements on silicon (Si) and polyimide (PI) substrates for solid-source and H$_2$Se recipes, while **Fig. 3c** illustrates the average and standard deviation for Se and W values, as well as the Se:W ratio. For as-grown samples (on Si), there is a ~20% discrepancy between the compositions of solid-source and H$_2$Se recipes. This is mainly a measurement artifact caused by the proximity of W$_M$ and Si$_{K\alpha}$ peaks at 1.774 and 1.739 keV, respectively, which causes peak fitting errors when translating the EDS peaks to atomic concentration. We avoid the Si peak by transferring WSe$_2$ films to a PI substrate. The EDS data from WSe$_2$ on PI substrates indicate very similar film compositions from each recipe (**Fig. 3c**), with a discrepancy of ~1% between the solid-source and H$_2$Se recipes, within the detection limit of EDS (1 atomic %). We also observe high spatial homogeneity for each recipe, with a coefficient of variation below 3% for Se and W atomic percentages (see tight error bars in **Fig. 3c**).

A transmission electron microscopy cross-section (**Fig. 3d**) confirms the van der Waals layered nature of the as-grown SS-Se films. Comparing the interlayer distances with the WSe$_2$ structure in the ICSD database ($a = b = 3.28$ Å, $c = 12.96$ Å, $\alpha = \beta = 90°$, $\gamma = 120°$)[19] and considering the atomic numbers of the constituting elements, each bright line is a plane containing W atoms, sandwiched between two Se layers (see **Fig. 3d**). The EDS result illustrated in **Fig. 3f and 3g** shows the elemental distribution in the lamella. Strong W and Se signals are seen atop the silicon substrate, confirming the elemental distribution in the film.

2θ-ω x-ray diffraction (XRD) scans verify the crystallinity of the selenized films, and the full-width at half-maximum is used to estimate the grain size using the Scherrer equation[20]. The presence of only (00ℓ) out-of-plane peaks for SS-Se WSe$_2$ in **Fig. 3h** demonstrates that the films are layered with alternating van der Waals gaps and undetectable out-of-plane crystallographic orientations, as verified by TEM (**Fig. 3d**). Estimates based on Scherrer equation also indicate grain sizes (coherence lengths) approximating the film thickness (**Fig. 3e**) for the SS-Se growth (See **Supplementary Table 2**). On the other hand, H$_2$Se samples leave more room for improvement. While H$_2$Se films do show in-plane XRD peaks (**Fig. 3h**), further optimization of growth parameters (such as growth promoters, carrier gas, or pressure optimization) could provide more ideal film orientation, and larger grains, as the partial pressure of Se during growth can determine the film orientation[15].

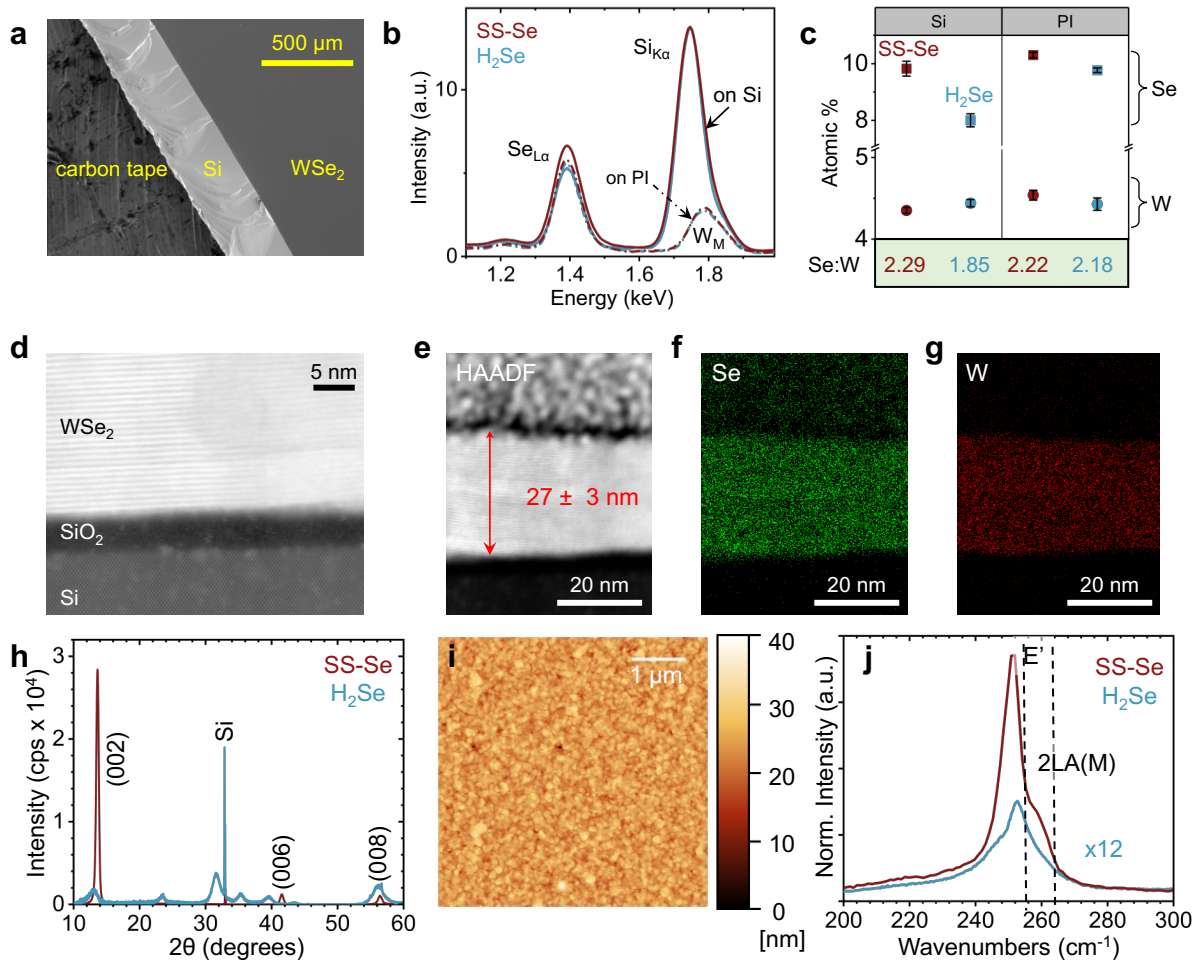

**Fig. 3. Material characterization of WSe$_2$. a,** H$_2$Se-grown films imaged under the scanning electron microscope (SEM) at the edge of the chip, **b,** resulting energy dispersive x-ray spectra of WSe$_2$ as-grown on Si and transferred onto polyimide (PI) in order to deconvolute the W emission from the Si substrate, **c,** resulting Se:W ratio of approximately 2:1, as anticipated for a fully-converted WSe$_2$ film. **d,** High angle annular dark field scanning transmission electron microscopy (HAADF-STEM) cross-section of WSe$_2$ samples from SS-Se growth. **e,** High-angle annular dark-field scanning TEM (HAADF-STEM) cross-section of the WSe$_2$ film with **f,** Se and **g,** W elemental mapping. The top layers are protective Pt/C coatings added during sample preparation. **h,** X-ray diffraction spectra of blanket films, measured on silicon. **i,** Film topography measured by atomic force microscopy **j,** Room temperature Raman spectra of representative films on Si.

We also display Raman spectra of both selenization methods in **Fig. 3j**, normalized to the Si peak of the substrate. These provide information about the vibrational modes, which correlate with the relative quality of the films[21]. Solid source grown films exhibit a distinct degenerate $A_1'$/E' peak (referred to as the E' peak for simplicity) and the 2LA(M) signature. $H_2Se$ films show an increased left shoulder to the E' peak, which could be due to a higher presence of defects in these films[22] or to degenerate peak splitting due to an increased presence of strain[23], as indicated by the shifted out of plane $(00\ell)$ peaks in the x-ray diffraction spectra (**Supplementary Table 2**).

Square $WSe_2$ samples measured by the van der Pauw method are used to determine the majority carrier type, density, and their Hall mobility (fabrication details in **Supplementary Note 2**). The films show p-type doping, with hole density ~$10^{17}$ cm$^3$ and average Hall mobility of ~5 cm$^{-2}$V$^{-1}$s$^{-1}$ at room temperature (**Supplementary Fig. 4**). Across various samples, we measured Hall mobilities and hole densities up to 8 cm$^2$ V$^{-1}$ s$^{-1}$ and $4.2 \times 10^{17}$ cm$^{-3}$, respectively. These are among the best mobility results reported to date for selenized $WSe_2$ (**Supplementary Table 2**). SS-Se samples exhibit a tight distribution of Hall mobility and hole density whereas $H_2Se$ samples show wider variation (**Supplementary Fig. 4**). This could be explained by the mobility anisotropy in $WSe_2$ and the more varied orientation of $WSe_2$ layers observed in the $H_2Se$ growths (**Supplementary Fig. 2**) causing sample-to-sample variation. In contrast, we note that SS-Se growths have better horizontal layering within the $WSe_2$ films.

## Flash-photolysis time-resolved microwave conductivity (TRMC) measurement

The flash-photolysis time-resolved microwave conductivity (TRMC) method is adopted to determine the charge carrier lifetime in our selenized films, particularly as this dictates the efficiency limit in photovoltaic applications[8]. TRMC is a contactless microwave spectroscopy technique where a thin film sample is excited by a laser inside a resonant cavity, probed by microwave radiation, and measured as a function of time[24]. TRMC experiments provide the product of free-charge carrier yield and the sum of both electron and hole mobilities, known as the yield-mobility product $\phi \sum \mu = \phi(\mu_e + \mu_h)$, where $\phi$ is the yield of charges per photon, while $\mu_e$ and $\mu_h$ are the electron and hole mobility, respectively. Both electrons and holes are produced by photoexcitation, but the degree to which carriers experience trap sites (and the precise energies of those traps) is unclear to date, in TMDs. Thus, it is challenging to deconvolve $\phi \sum \mu$ to estimate the photogenerated carrier mobilities. Instead, we focus on the carrier lifetimes and report the yield-mobility product (normalized by the fraction of absorbed photons) for three excitation wavelengths.

To characterize TMD photoconductivity, we first examine the optical absorption and steady-state microwave conductivity (SSMC) action spectra[24-30] associated with a TMD film for which we could measure a sufficiently strong response at low fluences (**Supplementary Fig. 5** and **Supplementary Fig. 6**). The optical absorption and SSMC action spectra provide the three excitation wavelengths (at 500, 580, and 780 nm) for the TRMC study, as follows: 1) bulk transitions above the WSe$_2$ band gap (500 nm or 2.48 eV), 2) transitions into the B exciton (580 nm or 2.14 eV), and 3) transitions into the lowest-lying A exciton (780 nm or 1.59 eV)[31]. **Figures 4a-c** display TRMC transients at varying fluences for the SS-Se films, at the three excitation wavelengths. **Figures 4d** and **4e** show the peak $\phi \sum \mu$ vs. fluence and the amplitude-weighted average carrier lifetimes vs. fluence, respectively. **Figure 4f** displays carrier lifetimes in WSe$_2$ films obtained from both SS-Se (red) and H$_2$Se (blue) growth methods at a fluence of $10^9$ cm$^{-2}$, representing AM 1.5 G solar illumination. (These lifetimes are extrapolated from the average plateaus at low fluence in **Fig. 4e**.)

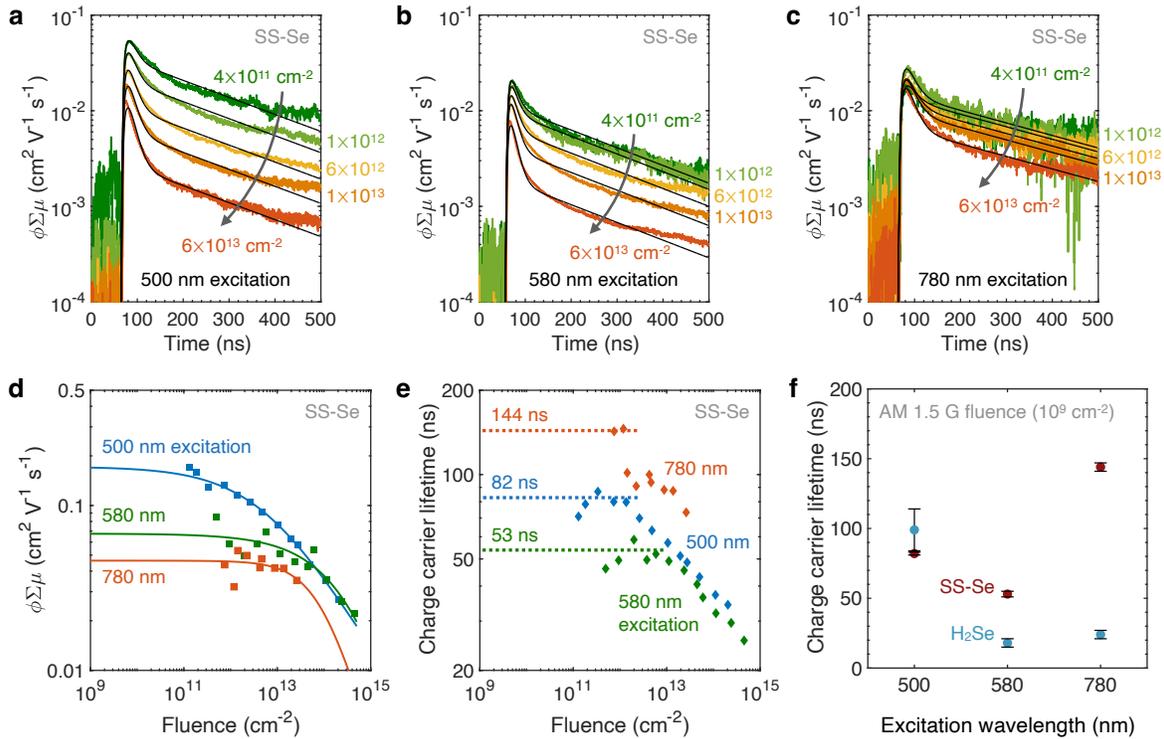

**Fig. 4. Flash-photolysis time-resolved microwave conductivity (TRMC) measurements.** TRMC transients of WSe$_2$ grown by SS-Se at varying fluences (from ≈ $10^{11}$ to $10^{14}$ cm$^{-2}$) for excitation wavelengths at **a**, 500 nm, **b**, 580 nm, and **c**, 780 nm. Some transients are omitted for clarity. Solid lines represent a double exponential fit to the series of transients. **d**, Peak yield-mobility product vs. fluence at three excitation wavelengths (500, 580, and 780 nm) for WSe$_2$ grown by SS-Se. Square symbols are measurements; solid lines show a Dicker-Ferguson fit[32, 33] to each data set, accounting for high-order recombination from exciton-exciton and exciton-charge quenching mechanisms. **e**, Amplitude-weighted average charge carrier lifetime vs. fluence at the same excitation wavelengths. Diamond symbols are measurements; dashed lines indicate charge carrier lifetimes averaged at low fluence, representing AM 1.5 G solar illumination ($10^9$ cm$^{-2}$). **f**, Charge carrier lifetimes under AM 1.5 G fluence vs. the three excitation wavelengths, for WSe$_2$ films prepared by SS-Se (symbols in Stanford Cardinal red) and H$_2$Se (imec blue) selenization. Error bars are approximated as standard deviation of carrier lifetimes in the flat, plateaued regions of panel **e**.

The data summarized in **Fig. 4d** and **4e** show that both the yield-mobility product and the carrier lifetimes depend on fluence for the 500 nm and 580 nm excitations, but the excitation at 780 nm yield less fluence-dependence. This is explained by the different initial excitation energies with respect to the WSe$_2$ band gap, i.e. the bulk and B exciton energies (2.48 and 2.14 eV, i.e. 500 and 580 nm) are greater than the WSe$_2$ optical band gap (1.59 eV or 780 nm) and have overall lower Coulomb binding energies. Higher energy excitations (580 nm vs. 700 nm, **Fig. 4d**) lead to a greater yield of free charges, under the assumption that the mobility is equal in both cases. This trend suggests either that the exciton binding energy decreases as the excitation energy increases, or that the relaxation process from higher energies to the minimum optical transition allows for more excitons to dissociate. However, the 500 nm excitations produce charges with a longer lifetime than 580 nm, despite their higher charge density. This unusual trend in lifetime could be due to defect sites facilitating charge separation, which are only made accessible by higher energy excitations, meaning that different types of defects could dominate the 500 nm vs. 580 nm charge separation and recombination.

As shown in **Fig. 4f**, SS-Se films tend to have higher carrier lifetimes than H$_2$Se samples, these results being the first of their kind in selenized WSe$_2$ films, to our knowledge. This could be related to the larger grain size in WSe$_2$ films grown by SS-Se (see **Supplementary Table 2**). The carrier lifetime of 144 ns measured in this work is over 14× higher than previous demonstrations of large-area TMD films (up to 10 ns in liquid exfoliated WS$_2$)[16]. Our large carrier lifetimes can be attributed to the vdW layered structure and large grain size of our selenized films. Carrier lifetimes demonstrated here are on the same order of magnitude as the highest value reported for multilayer TMDs (611 ns for WS$_2$ grown by chemical vapor transport[7]), as well as more established chalcogenides CdTe and CIGS[34, 35], confirming the high quality of our synthesis, particularly for photovoltaic applications.

## Projected efficiency and specific power of WSe$_2$ solar cells and modules

The ultimate performance limit of a single-junction solar cell is governed by the optoelectronic characteristics and synthesis quality of its absorber material. We perform a realistic modeling of single-junction solar cells performance limits with WSe$_2$ absorber layers by accounting for both intrinsic and extrinsic properties (**Fig. 5**). The model incorporates measured absorption data as well as radiative, Auger, and defect-assisted Shockley-Read-Hall (SRH) recombination, as described in detail in our previous work[8].

**Figure 5a** shows the simulated current density vs. voltage and power conversion efficiency (PCE) limit of single-junction solar cells with 26 nm-thick WSe$_2$ absorber layers at various film quality levels, as represented by SRH recombination lifetimes ($\tau_{SRH}$). This thickness is chosen here because it is that of our selenized films, and is the thickness which maximizes PCE[8] for lifetimes in the 10 ns to 1 µs range. Our model shows that $\tau_{SRH}$ of 10 ns, 100 ns, and 1 µs lead to PCE limits of 18.2%, 21.7%, and 25.1%, respectively (**Fig. 5a**). In the absence of SRH recombination (defect-free WSe$_2$, $\tau_{SRH} = \infty$), these 26 nm thin WSe$_2$ solar cells reach a PCE of 29.5% at the Tiedje-Yablonovitch limit (or up to ~30.8% in films thicker than 100 nm[8]). Carrier lifetimes associated with radiative and Auger recombination in intrinsic or lightly-doped multilayer WSe$_2$ are on the order of tens of micro-seconds[8]. Thus, measured carrier lifetimes below 1 µs correspond to SRH dominating recombination, which is the case for our selenized WSe$_2$ films (**Fig. 4f**). According to our realistic model, the selenized WSe$_2$ films in this study, which exhibit $\tau_{SRH}$ up to 144 ns, can achieve a PCE of 22.3% (**Supplementary Table 3**) upon design optimization, which is on par with incumbent solar technologies such as Si, CdTe, and CIGS[5].

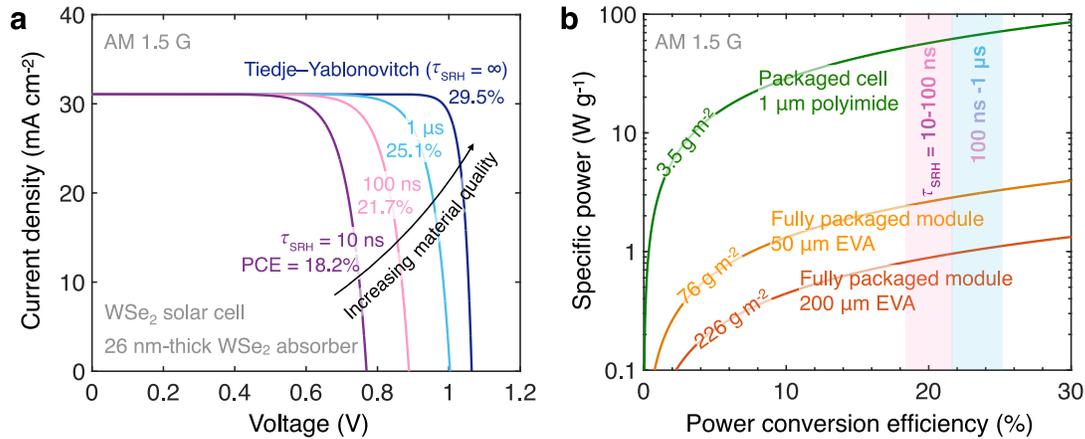

**Fig. 5. Projected performance of WSe$_2$ solar cells and solar modules. a,** Simulated current density-voltage characteristics and power conversion efficiency (PCE) limits of 26 nm-thick WSe$_2$ solar cells at various synthesis quality levels, represented by SRH recombination lifetime ($\tau_{SRH}$). Our selenized WSe$_2$ films have carrier lifetime up to 144 ns, corresponding to a PCE of 22.3% in an optimized design (**Supplementary Table 3**). **b,** Projected specific power of WSe$_2$ solar cells and modules. Upon integration on ultrathin flexible substrates, selenized WSe$_2$ solar cells can provide specific power of ~64 W g$^{-1}$ in a packaged cell, over 10× higher than established thin-film technologies[10]. Fully-packaged modules, which account for interconnect weight and EVA (ethylene-vinyl acetate) encapsulation[5], show specific power up to 3 W g$^{-1}$, over 4× higher than present record of 0.7 W g$^{-1}$ in a multi-junction III-V based solar module[5].

More notably, after integration on ultrathin flexible substrates, selenized WSe$_2$ solar cells are expected to provide specific power of ~64 W g$^{-1}$ in a packaged cell (**Fig. 5b**), over 10× higher than established thin-film solar cell technologies such as III-Vs, CdTe and CIGS[10]. These packaged WSe$_2$ solar cells could be adopted in size-constrained, low-power applications such as Internet of Things (IoT) and wearable electronics. Higher-power applications such as satellites and electric vehicles require larger, fully-packaged modules, with higher areal weight densities due to additional interconnects and module encapsulation[5]. Such fully-packaged WSe$_2$ modules have specific power of up to 3 W g$^{-1}$, over 4× higher than the present record of 0.7 W g$^{-1}$ in multi-junction III-V solar modules[5].

## Conclusion

We report a scalable method for large-area, high-quality WSe$_2$ film synthesis through both solid-source and gaseous precursors. The films show a vdW layered structure and grain sizes on the order of the film thickness. We report some of the highest Hall mobilities to date, up to 8 cm$^2$ V$^{-1}$ s$^{-1}$, in uniform, wafer-scale, multilayer films. These films have charge carrier lifetimes up to 144 ns, corresponding to a solar cell efficiency of ~22.3%, as well as specific power of ~64 W g$^{-1}$ in a packaged cell and 3 W g$^{-1}$ in a fully-packaged module. These results pave the way for large-scale, commercial TMD solar cells with high specific power, which could address the growing need for lightweight, flexible energy harvesters across various sectors such as IoT, wearables, electric vehicles, and aerospace.

## Methods

**Growth.** Solid source selenium selenization was carried out in a two-zone PlanarTech CVD system, with the primary zone consisting of a hatch furnace and the second heating zone of an insulating jacket with heating coils in order to vaporize the solid source selenium.

Gas source selenization was done in a cold-wall rapid thermal processing furnace (Annealsys AS-One 150), capable of housing 150 mm wafers. The heat is provided by infrared halogen lamps on the chamber's lid, and the temperature is monitored using a thermocouple attached to the susceptor. Processing gases were nitrogen, and 10% diluted H$_2$Se in N$_2$.

**TEM.** To study the cross-section of the samples, a FIB lamella was prepared from each sample on a Cu Omniprobe TEM grid, using a Thermo Fisher Helios FIB-SEM and for characterization. High Angle Annular Dark Field Scanning Transmission Electron Microscopy (HAADF-STEM) and Energy Dispersive X-ray spectroscopy (EDS) were performed on an aberration-corrected Thermo Fisher Titan transmission electron microscope at 300 kV, using a Super X detector. To prevent damage from the ion beam on the sample, the surface was covered with a carbon-platinum layer and afterward coated with platinum.

**Raman**. Raman spectroscopy was taken using a HORIBA LabrRAM HR Evolution spectrometer using a laser wavelength of $\lambda$ = 532 nm at $P$ = 0.5 mW and at 1800 gr/mm, unless otherwise specified.

**AFM**. Atomic force microscopy is taken with a Park NX-10, using an NSC-15 tip at scan rate of 0.75 Hz.

**XRD**. X-Ray Diffraction is carried out with a PANAnalytical X'Pert 2 Diffractometer with a Cu-K$\alpha$ source. Samples are first aligned to the (400) silicon peak to ensure orientation in the $z$ direction, then measured using Cu-K$\alpha$ radiation through a ½" slit without a parallel plate collimator.

**TRMC**. The detailed process for producing a quantitative measurement from TRMC is given in detail in our previous work, Reid *et al* [24].

**Specific power calculation**: The specific power is calculated based on the solar cell efficiency and the areal weight densities of the solar cell or the module. The output power is equal to efficiency multiplied by the incident power of 100 mW cm$^{-2}$ (AM 1.5 G one-sun illumination), and specific power is equal to the output power divided by the areal density. The areal density of the packaged cell is calculated by summing up the areal densities of all materials in the packaged solar cell stack by using the volumetric mass density multiplied by the respective material thickness, as explained in detail in our previous study[10]. A material stack of 1 μm polyimide (substrate), 80 nm gold (back contact/reflector), 26 nm WSe$_2$ (absorber), single-layer graphene (top contact), and 70 nm MoO$_x$ (anti-reflection coating/encapsulation layer) is considered as an example device structure[10]. The areal densities of the packaged modules are taken from the literature[5]. The values correspond to a lightweight, all-plastic packaging compatible with moisture-insensitive, thin-film solar technologies.

## Contributions

K.M.N., S.H., and K.N. contributed equally. K.M.N., S.H., K.N. and A.D. conceived the project. K.M.N. and S.H. performed the growth of WSe$_2$. S.H. carried out the SEM and EDS measurements on WSe$_2$ films. S.R. performed the TEM and subsequent EDS measurements. K.M.N. performed Raman, XRD, and AFM characterization of the films. K.N. fabricated the devices and conducted the Hall measurements. K.N. and F.U.N. modeled the solar cell performance. K.N. did the specific power calculations. J.C., J.B. and O.R. conducted the TRMC measurements. K.M.N., S.H., and K.N. wrote the manuscript, with assistance from A.D., J.C., and S.R. E.P. supervised the work. All authors contributed to the data interpretation, presentation, and revision of the manuscript.

## Acknowledgements

K.M.N. acknowledges support from the Stanford Graduate Fellowship (SGF) and from the National Science Foundation Graduate Research Fellowship (NSF-GRFP). S.H. acknowledges financial support by the Flanders Research Foundation (FWO)—strategic basic research doctoral grant 1S31922N. K.N. acknowledges partial support from Stanford Precourt Institute for Energy and the member companies of the SystemX Alliance at Stanford. A.D. acknowledges support by the Deutsche Forschungsgemeinschaft (DFG, German Research Foundation) through the Emmy Noether Programme (506140715). Part of this work was performed at the Stanford Nano Shared Facilities (SNSF), supported by the National Science Foundation under award ECCS-2026822. Additionally, this work was authored, in part, by the National Renewable Energy Laboratory, operated by Alliance for Sustainable Energy, LLC, for the U.S. Department of Energy (DOE) under Contract No. DE-AC36-08GO28308. Microwave conductivity measurements and analysis at NREL was funded by the Solar Photochemistry Program, Division of Chemical Sciences, Geosciences, and Biosciences, Office of Basic Energy Sciences, U.S. DOE. The views expressed in the article do not necessarily represent the views of the DOE or the U.S. Government.

**Author Declaration:** The authors declare no competing interests.

**Data Availability:** The data that support the findings of this study are available from the corresponding author upon reasonable request.# References

1. Das S, Sebastian A, Pop E, McClellan CJ, Franklin AD, Grasser T, Knobloch T, Illarionov Y, Penumatcha AV, Appenzeller J, Chen Z, Zhu W, Asselberghs I, Li L-J, Avci UE, Bhat N, Anthopoulos TD, Singh R. Transistors based on two-dimensional materials for future integrated circuits. *Nature Electronics* 2021, **4**(11)**:** 786-799.

2. Jariwala D, Davoyan AR, Wong J, Atwater HA. Van der Waals Materials for Atomically-Thin Photovoltaics: Promise and Outlook. *ACS Photonics* 2017, **4**(12)**:** 2962-2970.

3. Daus A, Vaziri S, Chen V, Köroğlu Ç, Grady RW, Bailey CS, Lee HR, Schauble K, Brenner K, Pop E. High-performance flexible nanoscale transistors based on transition metal dichalcogenides. *Nature Electronics* 2021, **4**(7)**:** 495-501.

4. Kim K-H, Andreev M, Choi S, Shim J, Ahn H, Lynch J, Lee T, Lee J, Nazif KN, Kumar A, Kumar P, Choo H, Jariwala D, Saraswat KC, Park J-H. High-Efficiency $WSe_2$ Photovoltaic Devices with Electron-Selective Contacts. *ACS Nano* 2022, **16**(6)**:** 8827-8836.

5. Reese MO, Glynn S, Kempe MD, McGott DL, Dabney MS, Barnes TM, Booth S, Feldman D, Haegel NM. Increasing markets and decreasing package weight for high-specific-power photovoltaics. *Nature Energy* 2018, **3**(11)**:** 1002-1012.

6. Hu Z, Lin D, Lynch J, Xu K, Jariwala D. How good can 2D excitonic solar cells be? *Device* 2023, **1**(1)**:** 100003.

7. Went CM, Wong J, Jahelka PR, Kelzenberg M, Biswas S, Hunt MS, Carbone A, Atwater HA. A new metal transfer process for van der Waals contacts to vertical Schottky-junction transition metal dichalcogenide photovoltaics. *Science Advances*, **5**(12)**:** eaax6061.

8. Nassiri Nazif K, Nitta FU, Daus A, Saraswat KC, Pop E. Efficiency limit of transition metal dichalcogenide solar cells. *Communications Physics* 2023, **6**(1)**:** 367.

9. Nassiri Nazif K. Transition Metal Dichalcogenides for Next-Generation Photovoltaics. *Stanford University* 2021.

10. Nassiri Nazif K, Daus A, Hong J, Lee N, Vaziri S, Kumar A, Nitta F, Chen ME, Kananian S, Islam R, Kim K-H, Park J-H, Poon ASY, Brongersma ML, Pop E, Saraswat KC. High-specific-power flexible transition metal dichalcogenide solar cells. *Nature Communications* 2021, **12**(1)**:** 7034.

11. Nakamura M, Yamaguchi K, Kimoto Y, Yasaki Y, Kato T, Sugimoto H. Cd-Free Cu(In,Ga)(Se,S)$_2$ Thin-Film Solar Cell With Record Efficiency of 23.35%. *IEEE Journal of Photovoltaics* 2019, **9**(6)**:** 1863-1867.
13

# Supplementary Information

## Toward Mass-Production of Transition Metal Dichalcogenide Solar Cells: Scalable Growth of Photovoltaic-Grade Multilayer WSe$_2$ by Tungsten Selenization

Kathryn M. Neilson,[1,†] Sarallah Hamtaei,[2,3,4,†] Koosha Nassiri Nazif,[1,†] Joshua M. Carr,[5] Sepideh Rahimi,[6] Frederick U. Nitta,[1,7] Guy Brammertz,[2,3,4] Jeffrey L. Blackburn,[8] Joke Hadermann,[6] Obadiah G. Reid,[8,9] Bart Vermang,[2,3,4] Alwin Daus,[1,10] & Eric Pop[1,7,11]

[1]Dept. of Electrical Engineering, Stanford Univ., Stanford, CA 94305, USA

[2]Hasselt Univ., imo-imomec, Martelarenlaan 42, Hasselt 3500, Belgium

[3]Imec, imo-imomec, Thor Park 8320, Genk 3600, Belgium

[4]EnergyVille, imo-imomec, Thor Park 8320, Genk 3600, Belgium

[5]Univ. Colorado Boulder, Materials Science & Engineering Program, Boulder CO 80303, USA

[6]Univ. Antwerp, Electron Microscopy for Materials Science (EMAT), Groenenborgerlaan 171, Antwerpen 2020, Belgium

[7]Dept. of Materials Science & Engineering, Stanford Univ., Stanford, CA 94305, USA

[8]National Renewable Energy Laboratory, Chemistry and Nanoscience Center, Golden CO 80401, USA

[9]University of Colorado Boulder, Renewable and Sustainable Energy Institute, Boulder CO 80303, USA

[10]Sensors Laboratory, Dept. Microsystems Engineering (IMTEK), Univ. Freiburg, Freiburg, Germany

[11]Precourt Institute for Energy, Stanford Univ., Stanford, CA 94305, USA

[†]These authors contributed equally. *Corresponding author email: epop@stanford.edu

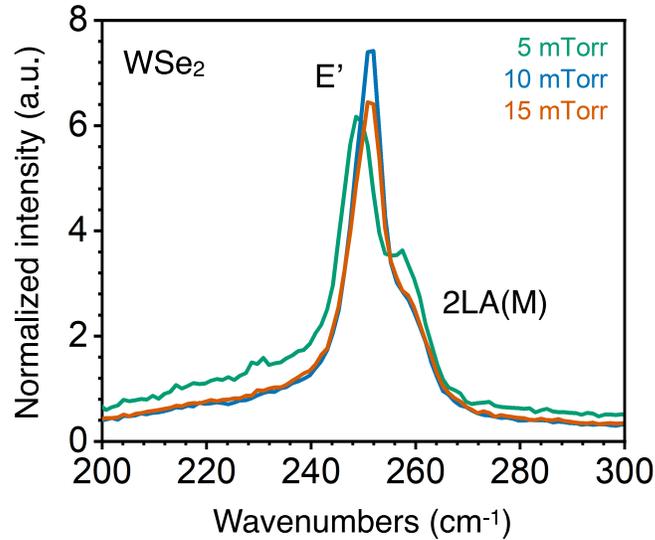

**Supplementary Fig. 1: Finding optimal W sputtering pressure via Raman spectroscopy.** The figure shows Raman spectra of as-grown $WSe_2$ with the starting W film sputtered at varying pressures. In order to optimize the W sputtering pressure, 10 nm films of W were sputtered at 5, 10, and 15 mTorr and subsequently grown using the same recipe as described in the main text. To determine an optimal W sputtering condition, we selenize W films deposited at three sputtering pressures and compare the intensity of the E'/$A_1$' peak to the 2LA(M) signature in $WSe_2$. The 2LA(M) vibrational mode is thought to become active due to structure disorder[21], therefore we compare the defect-related signature intensity to that of the E' and $A_1$' phonon mode. From this, it can be seen that a pressure of 10 mTorr gives the highest E'/$A_1$' intensity. Increasing sputtering pressures generally allow for more porosity in the film, but too much porosity could lead to a non-homogeneous film. Therefore, we believe that 10 mTorr could act as a balance between these two competing factors.

**Supplementary Table 1: Full analysis of films' elemental composition by energy dispersive x-ray spectroscopy (EDS).** In each sample, the mean and coefficient of variation (CV) is reported for three different measurements. Note that there is a carbon peak reported in all cases, which is a well-known contaminant in EDS analysis; any residual organic gases in the measurement chamber (or with the sample itself) is exposed to, and cracked by, the electron beam, and subsequently reflected in the response as carbon peak. As expected, this is further intensified for samples transferred onto PI. The source of oxygen can also be attributed to leftover oxygen in the chamber, or native oxides.

| Sample | | Statistics | Atomic concentration (%) | | | | | Ratio |
|---|---|---|---|---|---|---|---|---|
| | | | C | O | Si | Se | W | Se/W |
| On silicon | SS-Se | Mean | 5.24 | 3.39 | 76.88 | **10.10** | **4.39** | **2.29** |
| | | CV (%) | 1.31 | 10.04 | 0.66 | **2.60** | **0.85** | **1.93** |
| | $H_2Se$ | Mean | 4.54 | 1.61 | 81.14 | **8.26** | **4.45** | **1.85** |
| | | CV (%) | 3.23 | 9.85 | 0.54 | **2.79** | **1.01** | **2.84** |
| On polyimide | SS-Se | Mean | 70.22 | 14.97 | - | **10.23** | **4.59** | **2.22** |
| | | CV (%) | 1.08 | 5.27 | - | **0.95** | **1.32** | **2.26** |
| | $H_2Se$ | Mean | 70.85 | 14.91 | - | **9.77** | **4.47** | **2.18** |
| | | CV (%) | 1.13 | 6.39 | - | **0.84** | **1.71** | **1.08** |

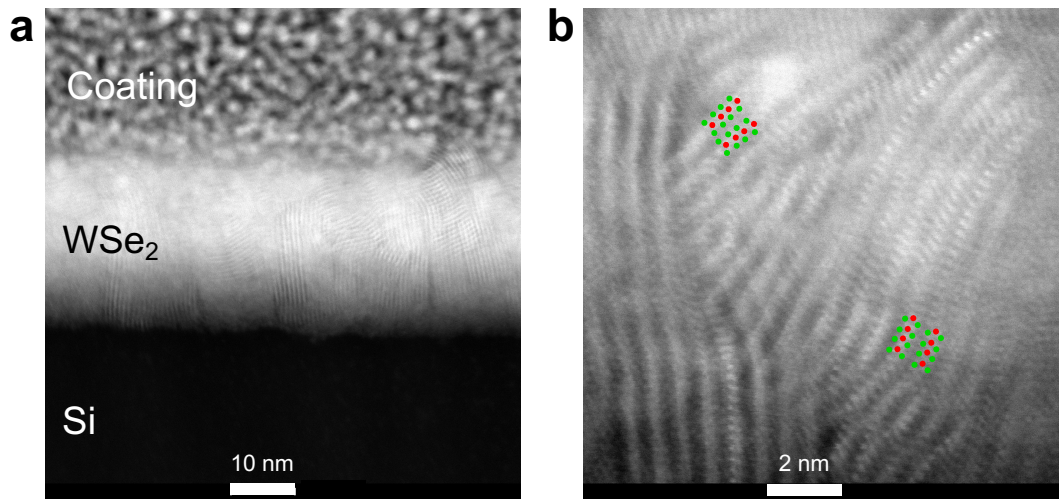

**Supplementary Fig. 2: Transmission electron microscopy (TEM) of H$_2$Se-selenized films. a,** Cross-sectional view of selenized films on a silicon substrate with a metallic coating. **b)** A higher magnification image showing the varied orientation of lamella. The H$_2$Se-selenized films are prepared as described in Methods section within the main text. Films show varied orientations, unlike the exclusively (00ℓ) orientation from solid source grown films shown in **Fig. 3d**. This corroborates the orientations seen in X-ray diffraction measurement (**Fig. 4a**). Further optimization of the growth parameters could further reduce the vertical orientation[36].

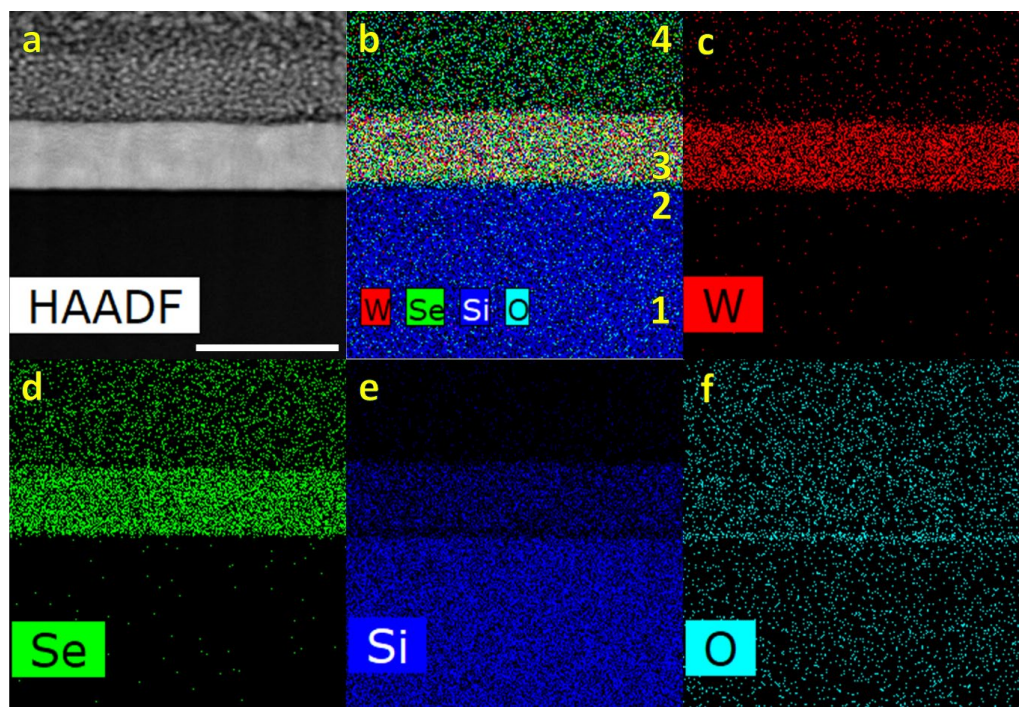

**Supplementary Fig. 3: Elemental mapping of H₂Se-selenized films using Energy Dispersive X-Ray Spectroscopy. a**, HAADF-STEM image. Scale bar, 60 nm **b**, the overlap of the elemental maps. Regions 1, 2, 3, and 4 refer to Si, SiO$_2$, WSe$_2$, and Pt-C protective layer, respectively. **c-f**, The elemental distribution of **c**, W, **d**, Se, **e**, Si, and **f**, O, respectively. All images are in the same scale.

**Supplementary Note 1. Method of extracting grain sizes from X-ray diffraction measurements**

Grain sizes and micro strain analysis are estimated using the Scherrer equation (**Supplementary Equation 1**) by fitting a Gaussian curve to the (002) diffraction peak in both solid source and H$_2$Se-selenized films. Subsequent estimated grain sizes can be found in **Supplementary Table 2.**

$$D = \frac{K \cdot \lambda}{FWHM \cdot \cos(\theta)} \quad (1)$$

Where:

    D = Crystallite/ grain size

    K = Scherrer shape factor, typically around 0.9[20]

    λ = Cu-Kα wavelength (1.54 Å)

    FWHM = Full width at half maximum at the diffraction peak

    θ = the Bragg angle of diffraction at a given peak

**Supplementary Table 2: Fit and calculated parameters from x-ray diffraction analysis.**

| Film | 2θ (°) | FWHM (°) | Grain Size (nm) | Strain |
|---|---|---|---|---|
| H$_2$Se | 13.16 | 0.978 | 8 | 3.5% |
| SS-Se | 13.62 | 0.478 | 17 | 0.02% |

Strain is calculated using lattice spacing reported in the ICSD database, estimated as follows, where *d* is the calculated lattice spacing from XRD and $d_0$ is the theoretical lattice spacing (**Supplementary Equation 2**). The theoretical lattice spacing is taken from the ICSD database[19] to estimate strain, although further un-idealities of the measurement can be included more a more precise value.

$$Strain\ (\%) = \frac{d - d_0}{d_0} \cdot 100 \quad (2)$$

**Supplementary Note 2. Method of extracting Hall mobility in selenized WSe$_2$ films**

Square WSe$_2$ samples having four contacts are prepared for the van der Pauw method to determine the majority carrier type, obtain the Hall mobility and density of majority carrier in the films using a DC Hall effect measurement System (Lakeshore 8404). The van der Pauw method is preferred as other field-effect type mobility measurements can suffer from inaccurate extractions due to an underestimated threshold voltage and contact resistance[37], both of which require further optimization. The measurement is performed on ~26 nm thick WSe$_2$ films transferred onto flexible insulating ~5 μm-thick polyimide substrates (**Supplementary Fig. 4a**). Four 2 mm x 2 mm gold (Au) contact pads are e-beam evaporated on top of the WSe$_2$ film in a square van der Pauw pattern with a side length of 10 mm (**Supplementary Fig. 4b**), which according to the tool's manual results in a measurement error less than 2%.

The films show p-type doping, and an average Hall mobility of ~5 cm$^{-2}$V$^{-1}$s$^{-1}$ and hole density of ~ $10^{17}$ cm$^3$ (**Supplementary Fig. 4c-d**). Hall mobilities and hole mobilities as high as 8 cm$^2$ V$^{-1}$ s$^{-1}$ and 4.2 x $10^{17}$ cm$^{-3}$ are measured, respectively. These values are among the best mobility results reported to date for selenized WSe$_2$ (**Supplementary Table 2**). SS-Se samples exhibit a tight distribution of hall mobility and hole density whereas H$_2$Se samples show wider variation. This could be explained by the mobility anisotropy in WSe$_2$ and the random orientation of WSe$_2$ layers observed the H$_2$Se growth (**Supplementary Fig. 2**), leading to sample-to-sample variation, as opposed to SS-Se growth, which mostly consists of horizontally oriented WSe$_2$ layers exhibiting in-plane mobility.

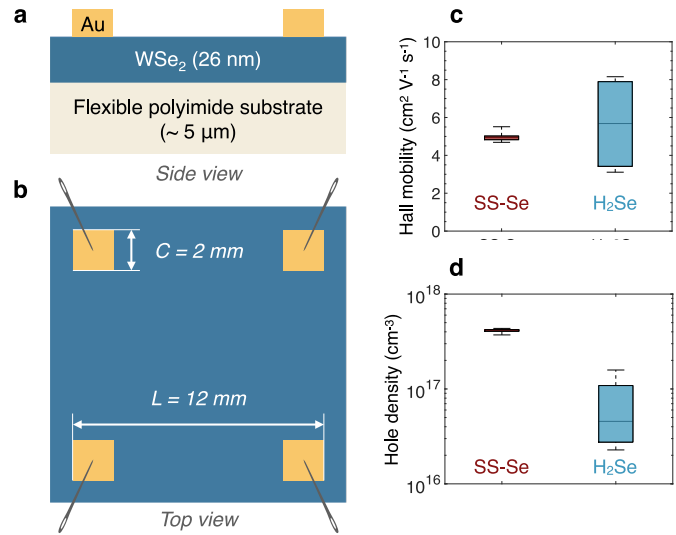

**Supplementary Fig. 4: Van der Pauw measurements to determine Hall mobility, majority carrier type and majority carrier density of WSe$_2$ films. a,** side view and **b,** top view of the square four-contact van der Pauw structure used to measure the Hall mobility, majority carrier type (i.e., holes) and hole density. **c,** Hall mobility and **d,** hole density of WSe$_2$ grown by selenization of tungsten in solid-source selenium (SS-Se) and H$_2$Se furnaces. Hall mobilities as high as 8 cm$^2$ V$^{-1}$ s$^{-1}$ are demonstrated. A maximum hole density of 4.2 x $10^{17}$ cm$^{-3}$ is measured in these films.

**Supplementary Table 2. Our selenization growth methods vs. state-of-the-art**

| Study | Growth method | WSe$_2$ thickness | Mobility (cm$^2$ V$^{-1}$ s$^{-1}$) | Mobility measurement method | Grain size (nm) | Film uniform? | Carrier lifetime | Growth temperature (°C) | Growth duration (hr) |
|---|---|---|---|---|---|---|---|---|---|
| **SS-Se (this work)** | W selenization | 22 nm | Up to 5 | Hall Effect | 17 | Yes | Up to 144 ns | <900 | >2 |
| **H$_2$Se (this work)** | W selenization | 22 nm | Up to 8 | Hall Effect | 8 | Yes | Up to 144 ns | <650 | >1 |
| Campbell et. al[38] | W/WO$_3$ selenization | 2.5 nm | 10 | Field effect (V$_T$ improperly extracted) | 400 | No | — | <800 | >1.5 |
| Kang et. al[12] | W selenization | 6 nm | 0.15 [FE] (V$_{ds}$ = –5 V) 22.3 [Hall] | Field Effect & Hall effect | — | Yes | — | <600 | — |
| Jäger-Waldau et. al[15] | W selenization on quartz | 400 nm to 4 μm | — | — | 10-35 | No | — | <850 | >18 |
| Li et. al[14] | W selenization | 25-200 nm | — | — | — | No | — | <600 | >1 |
| Hussain et. al[13] | W selenization | — | — | — | — | No | — | <600 | >1 |
| Medina et. al[39] | WO$_x$ selenization | 5 nm | 6 | Hall effect | — | Yes | — | <500 | >2 |
| Ma et. al[40] | W selenization | 100 nm (W) | 30 | Hall effect | 80 | No | — | <900 | >20 |
| Salitra et. al[41] | WO$_3$ selenization | 2 μm (WO$_3$) | — | — | 10-100 | No | — | <950 | >1 |
| Browning et. al[42] | WO$_3$ selenization | <10 nm | — | — | 50-200 | Yes | — | <900 | — |

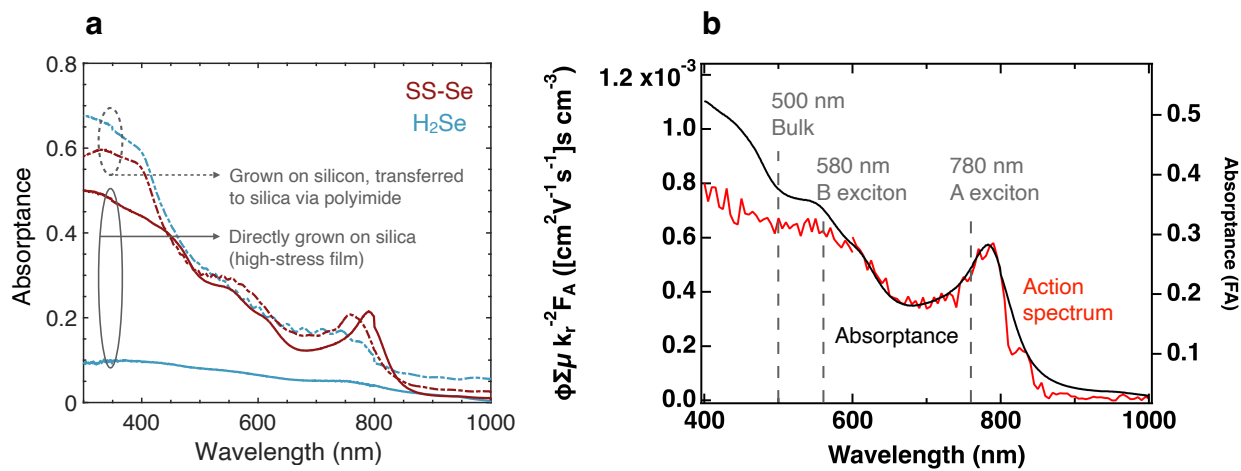

**Supplementary Fig. 5: Optical absorptance and steady-state microwave conductivity (SSMC) measurements to determine excitation wavelengths used in Flash-photolysis time-resolved microwave conductivity (TRMC). a,** Optical absorptance of SS-Se and $H_2Se$ $WSe_2$ films used in this study collected in a center-mounted configuration to account for reflection and transmission components simultaneously. **b,** SSMC action spectrum and optical absorption spectrum of SS-Se films. By first examining the optical absorption and action spectrum we can determine at what excitation wavelengths the TRMC experiments can be conducted to look for differences in the physical processes present in the optical absorption.

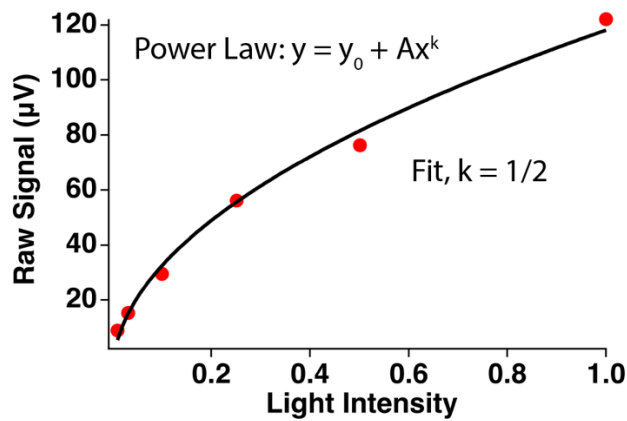

**Supplementary Fig. 6: Raw steady state microwave conductivity (SSMC) versus light intensity.** Plot of raw signal from the SSMC measurement for the SS-Se films versus incident light intensity. The light intensity is attenuated by addition of neutral density filters just before the entrance slit into the microwave cavity. Fitting the raw signal curve as a function of light intensity with a power law provides a molecularity relationship that demonstrates a square-root dependence, indicative of a bimolecular (second-order) recombination process which dominates the SSMC signal in the above figure. Fitting the raw signal curve as a function of light intensity with a power law provides a molecularity relationship that demonstrates a square-root dependence, indicative of a bimolecular (second-order) recombination process which dominates the SSMC signal in the above figure. Assuming that the excitation density in the SSMC is not approaching the trap density in these experiments, this observation is consistent with an SRH recombination dominated process with respect to the recombination loss. This molecularity fit is used in the SSMC analysis to ensure that the conductivity units for the y-axis are correct, particularly that the radiative recombination rate constant, $k_r$, is appropriately accounted for in the raw signal for calculating the action spectrum.

**Supplementary Table 3: Practical performance limits of 26 nm-thick WSe$_2$ solar cells at various material quality levels, as represented by Shockley-Reed-Hall (SRH) lifetimes, $\tau_{SRH}$.** The limits are calculated using a detailed balance model developed by Nassiri Nazif et al.[8], which incorporates measured optical absorption spectra and includes radiative, Auger, and defect-assisted SRH recombination mechanisms. WSe$_2$ solar cells made from our selenized WSe$_2$ films, which have $\tau_{SRH}$ of up to 144 ns, can in practice achieve 22.3% power conversion efficiency upon design optimization. Short-circuit current density; $V_{OC}$, open-circuit voltage; FF, fill factor; PCE, power conversion efficiency, MPP, maximum power point.

| $\tau_{SRH}$ | 10 ns | 100 ns | 144 ns | 1 μs | ∞ (no SRH recombination) |
|---|---|---|---|---|---|
| $J_{SC}$ (mA cm$^{-2}$) | 31.0 | 31.0 | 31.0 | 31.0 | 31.0 |
| $V_{OC}$ (V) | 771 | 890 | 909 | 1005 | 1065 |
| Fill factor | 0.76 | 0.79 | 0.79 | 0.81 | 0.89 |
| $V_{MPP}$ (mV) | 637 | 749 | 766 | 860 | 976 |
| $J_{MPP}$ (mA cm$^{-2}$) | 28.7 | 29.0 | 29.0 | 29.2 | 30.2 |
| **PCE (%)** | **18.2** | **21.7** | **22.3** | **25.1** | **29.5** |

**Additional References**